\documentclass[preprint,pre,showkeys,amsmath]{revtex4}
\usepackage{epsfig}

\newcommand{\unbfm}[1]{\mbox{\bf #1}}            

\begin{document} 

\title{
Decomposition of multicomponent mass spectra using Bayesian probability theory}
\author{
H.D. Kang}
\affiliation{Centre for Interdisciplinary Plasma Science, Max-Planck-Institut f\"{u}r Plasmaphysik, EURATOM Association, D-85748 Garching b. M\"{u}nchen, Germany}
\author{R. Preuss}
\affiliation{Centre for Interdisciplinary Plasma Science, Max-Planck-Institut f\"{u}r Plasmaphysik, EURATOM Association, D-85748 Garching b. M\"{u}nchen, Germany}
\author{T. Schwarz-Selinger}
\affiliation{Centre for Interdisciplinary Plasma Science, Max-Planck-Institut f\"{u}r Plasmaphysik, EURATOM Association, D-85748 Garching b. M\"{u}nchen, Germany}
\author{V. Dose}
\affiliation{Centre for Interdisciplinary Plasma Science, Max-Planck-Institut f\"{u}r Plasmaphysik, EURATOM Association, D-85748 Garching b. M\"{u}nchen, Germany}

\date{\today}

\begin{abstract}
We present a method for the decomposition of mass spectra of mixture gases using Bayesian probability theory. 
The method works without any calibration measurement and therefore applies also to the analysis of spectra containing unstable species. 
For the example of mixtures of three different hydrocarbon gases the algorithm provides concentrations and cracking coefficients of each mixture component as well as their confidence intervals.
The amount of information needed to obtain reliable results and its relation to the accuracy of our analysis are discussed.
\end{abstract}
\maketitle
\thispagestyle{empty}

\newpage
\section{Introduction}

Mass spectrometry is a standard technique for residual gas analysis and active processing control in vacuum devices.
Traditional low resolution quadrupole mass spectrometers are widely used due to high sensitivity, reasonable stability, wide operational pressure range, high scan speed and low costs.
To be filtered in the quadrupole field, neutral gases have to be ionized first, most commonly by electron impact. 
At a typical electron energy of 50-100 eV used to achieve a high ionization efficiency and stability analyte molecules
can decompose into a variety of fragment ions leading to the so called {\it cracking pattern}(CP).
The fragmentation is a molecular specific property which reflects the chemical structure and the atomic composition of the molecule. It can therefore be used for the decomposition of multicomponent mass spectra.

A direct decomposition of mass spectra from mixture gases by successive subtraction of component contributions from the mixture signal, the wide spread pedestrian approach, works only if there are non-interfering mass numbers.
A more elaborate method, {\it least square evaluation}, is not subject to this limitation and can incorporate measurement errors into the analysis\cite{dobrozemsky72,dobrozemsky92}. 
In both cases one must assume the CP of the components to be exactly known \cite{dobrozemsky72}.
The CP is, however, not an intrinsic property of molecules but depends on the particular mass spectrometer and operating parameters making the determination of CP nontrivial.
The problem becomes even more severe if unstable species like radicals are among the components of a mixture, since in this case there are neither calibration measurements nor literature values for the CP available. 
In case of mixtures containing a single radical the CP of the radical is usually identified with the rest signal after subtraction of all component signals from the mixture signal. 
Because of the error propagation of measured signals the above procedure, however, provides only poor and sometimes unphysical (e.g. negative cracking coefficients) estimations.

To overcome these difficulties and to treat measured spectra consistently we introduced a novel method using Bayesian probability theory \cite{schwarz01}.
This was applied to the pyrolysis of azo-methane in order to evaluate concentrations and improved cracking coefficients together with their confidence intervals for the pyrolysis products including the methyl radical.
In the present work we deal with another case of decomposition, in which mixtures made up given different compositions have to be decomposed without using any calibration measurements.
Because of lack of calibration measurements this method offers a practical use, in particular, in handling unknown mixture gases.
But most importantly, this approach is able to treat mixtures containing unstable species, because exact CPs are not needed.

As an example we synthesize mixtures of three different hydrocarbon gases, ethane(C$_2$H$_6$), propane(C$_3$H$_8$) and n-butane(C$_4$H$_{10}$).
The gas composition and the cracking coefficients of the mixture components will be estimated by independent experiments and compared with the results from the Bayesian analysis.

\section{Bayesian data analysis}

%
%

The procedure used in this work is similar to that used for the decomposition of the pyrolysis products of azo-methane introduced by Schwarz-Selinger {\it et al.} \cite{schwarz01}, where details of the calculation may be found.
Here we present only the relevant steps for our algorithm.

Assuming linear response of the mass spectrometer the mass signal $\tilde{\mathbf d}_j$ of a mixture is the sum of contributions of all species in the mixture
\
\begin{equation}
\tilde{\mathbf d}_j = \tilde{\mathbf C} \tilde{\mathbf x}_j + \boldsymbol{\varepsilon}_j
\quad ,
\end{equation}
\
where $j=1,...,J$ with $J$ mixtures.
The vector $\tilde{\mathbf d}_j$ has $N$ elements representing the same number of mass channels.
The composition vector $\tilde{\mathbf x}_j$ has the dimension $M$ of the number of species in the mixture.
$\tilde{\mathbf C}$ is a cracking matrix with $M$ column vectors corresponding to the cracking patterns of $M$ species.
Since we are interested in compositions and cracking patterns but not in the absolute intensities, $\tilde{\mathbf d}_j$ and the column vectors of $\tilde{\mathbf C}$ are normalized with respect to the sum of intensities of all mass channels for practical reasons, which enforces in turn the normalization of $\tilde{\mathbf x}_j$
\
\begin{equation}
\sum_{n=1}^N d_{nj} = \sum_{n=1}^N \sum_{m=1}^M \tilde{c}_{nm} \tilde{x}_{mj} = 
\sum_{m=1}^M \tilde{x}_{mj} \sum_{n=1}^N \tilde{c}_{nm} = 
\sum_{m=1}^M \tilde{x}_{mj} = 1
\quad ,
\end{equation}
\
where $\tilde{x}_{mj}$ is the {\it m}-th component of the {\it j}-th mixture.
This sum rule has been overlooked in our previous work \cite{schwarz01}.
The results in Ref.~3 approximate however closely the sum rule.
Quite the same it does improve stability of the estimates on $\tilde{\mathbf x}$ and $\tilde{\mathbf C}$ if the relation is explicitly included in the analysis.
The vector $\boldsymbol{\varepsilon}_j$ in Eq. (1) represents the finite noise of the experiments.
We assume that $\boldsymbol{\varepsilon}_j$ is bias free, $\langle \boldsymbol{\varepsilon}_j \rangle = 0$, and has a variance of $\langle \varepsilon^2_{nj} \rangle = \sigma^2_{nj}$.
Note that for the absolute composition of the mixtures $\tilde{\mathbf x}_j$ in (1) has to be scaled by a sensitivity factor of the mass spectrometer for molecules considered, which can be done by independent calibration experiments.
Since it is beyond the scope of this work it will not be discussed. In the following we call $\tilde{\mathbf x}_j$ concentration of a mixture.

Now we can make use of the normalization condition (2) and reduce the number of unknown concentrations by 1.
We split $\tilde{\mathbf C} \tilde{\mathbf x}_j$ in Eq. (1) into two parts
\
\begin{equation}
\tilde{\mathbf C} \tilde{\mathbf x}_j = \sum_{m=1}^M \tilde{\mathbf c}_m \tilde{x}_{mj} = \sum_{m=1}^{M-1} \tilde{\mathbf c}_m \tilde{x}_{mj} + \tilde{\mathbf c}_M \tilde{x}_{Mj}
\quad .
\end{equation}
\
Using $\tilde{x}_{Mj}=1-\sum_{m=1}^{M-1} \tilde{x}_{mj}$, $\tilde{\mathbf C} \tilde{\mathbf x}_j$ becomes
\
\begin{equation}
\tilde{\mathbf C} \tilde{\mathbf x}_j = \tilde{\mathbf c}_M + \sum_{m=1}^{M-1} (\tilde{\mathbf c}_m - \tilde{\mathbf c}_M) \tilde{x}_{mj}
\quad .
\end{equation}
\
Then Eq. (1) can be rewritten
\
\begin{equation}
\tilde{\mathbf d}_j - \tilde{\mathbf c}_M = \sum_{m=1}^{M-1} (\tilde{\mathbf c}_m - \tilde{\mathbf c}_M) \tilde{x}_{mj} + \boldsymbol{\varepsilon}_j
\quad .
\end{equation}
\
We define
\
\begin{eqnarray}
\mathbf d_j &=& \tilde{\mathbf d}_j - \tilde{\mathbf c}_M
\nonumber\\
\mathbf C &=& (\tilde{\mathbf c}_1 -\tilde{\mathbf c}_M , \tilde{\mathbf c}_2 -\tilde{\mathbf c}_M ,..., \tilde{\mathbf c}_{M-1} -\tilde{\mathbf c}_M)\\
\mathbf {x}_j^T &=& (\tilde{x}_{1j},\tilde{x}_{2j},...,\tilde{x}_{(M-1)j})
\nonumber
\quad ,
\end{eqnarray}
\
and obtain the simplified model equation
\
\begin{equation}
\mathbf d_j = \mathbf C \mathbf x_j + \boldsymbol{\varepsilon}_j
\quad .
\end{equation}
\
Knowing the model equation for the mass signal, which assumes $\langle \mathbf d_j - \mathbf C \mathbf x_j\rangle = 0$ and $\langle \left(\mathbf d_j - \mathbf C \mathbf x_j \right)_n^2 \rangle = \sigma_{nj}^2$, the principle of maximum entropy \cite{sivia96} leads to a Gaussian sampling distribution({\it likelihood}) for the data set of the $j$-th mixture
\ 
\begin{equation}
p ( \mathbf d_j | \mathbf x_j , \mathbf C , \mathbf S_j , I ) =
\frac{1}{\prod_n s_{nj} \sqrt{2 \pi}}
\exp \left\{ - \frac12
\left( \mathbf d_j - \mathbf C \mathbf x_j \right)^T
\mathbf S_j^{-2}
\left( \mathbf d_j - \mathbf C \mathbf x_j \right)
\right\}
\quad,
\end{equation}
\
where $\left( \mathbf S_j^{-2} \right)_{nn} = 1/s_{nj}^2$ and $\left( \mathbf S_j^{-2} \right)_{nl} = 0$ for $n\neq l$,
with $s_{nj}$ being the {\it n}-th component of the sample variance $\mathbf s_j$ used for $\mathbf \sigma_{nj}$.
$I$ denotes all background information available.
We assume independent measurements.
This means that the outcome of an experiment is not influenced by any other data set.
In such a case the likelihood for all data sets reduces to a product of the likelihoods of the individual measurements 
\
\begin{equation}
p \left(
\mathbf D | \mathbf X,\mathbf C,\mathbf S,I \right) 
=
\prod_j p \left( \mathbf d_j | \mathbf x_j , \mathbf C ,  \mathbf S_j , I \right)
\quad .
\end{equation}
\
This can easily be shown by repeated application of the product rule of the Bayesian probability theory $p(A,B)=p(A)\ p(B|A)$. 
For reasons of convenience we used the notation 
$ \left\{ \mathbf d_j \right\} \equiv \mathbf D$,
$ \left\{ \mathbf x_j \right\} \equiv \mathbf X$ and
$ \left\{ \mathbf S_j \right\} \equiv \mathbf S$,
where $\mathbf D$ and $\mathbf X$ are matrixes with $J$ column vectors $\mathbf d_j$ and $\mathbf x_j$, respectively.

%
%
Our task is to find the concentration ${\mathbf x}_j$ ($j=1,...,J$) and the cracking coefficients $\tilde{\mathbf c}_m$ ($m=1,...,M$). 
Let us start with $\mathbf x_j$.
The expectation value of $\mathbf x_k$ is derived from the probability density $p(\mathbf X|\mathbf D, \mathbf S, I)$ as
\
\begin{equation}
\langle \mathbf x_k \rangle =
\int d\mathbf X\ \mathbf x_k \ p(\mathbf X|\mathbf D, \mathbf S, I)
\quad .
\end{equation}
\
$p(\mathbf X|\mathbf D, \mathbf S, I)$ is obtained from Bayes theorem as \cite{sivia96}
\
\begin{equation}
p(\mathbf X|\mathbf D, \mathbf S, I)=
p(\mathbf X |I)\ p(\mathbf D |\mathbf X, \mathbf S, I)/Norm
\quad ,
\end{equation}
\
in terms of the marginal likelihood $p(\mathbf D |\mathbf X, \mathbf S, I)$.
The latter is derived by the marginalization rule 
\
\begin{equation}
p(\mathbf D |\mathbf X, \mathbf S, I)=
\int d\tilde{\mathbf C}\  p(\mathbf D, \tilde{\mathbf C}|\mathbf X, \mathbf S, I)=
\int d\tilde{\mathbf C}\  p(\tilde{\mathbf C}|I)\ p(\mathbf D|\tilde{\mathbf C}, \mathbf X, \mathbf S, I)
\quad .
\end{equation}
\
Note that $p(\tilde{\mathbf C}|\mathbf X, \mathbf S, I)=p(\tilde{\mathbf C}|I)$, since the composition and the sample variance do not contain any information about the cracking coefficients.
We use a flat prior for $p(\mathbf X |I)$ in the range of $0<x_{mj}<x_{max,j}$, since there exists no prior knowledge about the composition, and obtain 
\
\begin{equation}
\langle \mathbf x_k \rangle \sim
\int d\tilde{\mathbf C}\ p(\tilde{\mathbf C} |I) \int d\mathbf X\ \mathbf x_{k}\ p(\mathbf D |\mathbf X, \tilde{\mathbf C}, \mathbf S, I)
\quad ,
\end{equation}
\
where the proportionality accounts for the normalization factor in Eq. (11) and the flat prior $p(\mathbf X |I)$.
For the analytical integration of the inner integral in (13) we assume that the integrand is sharply peaked in the prior range for $\mathbf x_j$ so that the integral does not change if the integral limits are extended to $(-\infty,\infty)$.
The likelihood $p(\mathbf D |\mathbf X, \tilde{\mathbf C}, \mathbf S, I)$ in Eq. (13) is a multivariate Gaussian function (8,9), whose exponent can be transformed into a complete square
\
\begin{equation}
\left(\mathbf d_j - \mathbf C \mathbf x_j \right)^T
\mathbf S_j^{-2}
\left(\mathbf d_j - \mathbf C \mathbf x_j \right)
=
\left(\mathbf x_j - \mathbf x_{j0} \right)^T
\mathbf Q_j
\left(\mathbf x_j - \mathbf x_{j0} \right)
+ \mathbf R_j
\quad .
\end{equation}
\
The integral of the right hand side of Eq. (14) over $\mathbf x_j$ with range $(-\infty,\infty)$ returns 
\
\begin{equation}
\exp \left( - \frac{R_j}{2} \right) |\det\mathbf Q_j|^{-\frac12}
\quad ,
\end{equation}
\
and thus for the rightmost integral of (13) we obtain
\
\begin{equation}
\int d\mathbf X\ \mathbf x_{k}\ p(\mathbf D |\mathbf X, \tilde{\mathbf C}, \mathbf S, I) =
\mathbf x_{k0}\ \prod_{j} \exp ( - \frac{R_j}{2} ) | \det \mathbf Q_j |^{-\frac12}
\quad ,
\end{equation}
\
where $\mathbf x_{k0}$, $R_j$ and $\mathbf Q_j$ result from the comparison of the coefficients in Eq. (14)
\
\begin{eqnarray}
\mathbf Q_j &=& \mathbf C_j^T \mathbf S_j^{-2} \mathbf C_j
\nonumber\\
\mathbf x_{k0} &=& \mathbf Q_k^{-1} \mathbf C_k^T \mathbf S_k^{-2} \mathbf d_k \\
R_j &=& \mathbf d_j^T \mathbf S_j^{-2} \mathbf d_j - \mathbf x_{j0}^T \mathbf Q_j \mathbf x_{j0}
\nonumber
\quad .
\end{eqnarray}
\
The calculation of the matrix equations can be considerably simplified by using {\it singular value decomposition} (SVD) of $\mathbf C_j$. For details readers are referred to Ref. 3 and 5.

%
%
To complete the integration (13) the prior probability $p(\tilde{\mathbf C}|I)$ for the cracking coefficients has to be assigned. 
As a prior knowledge $I$ we use the table of {\it Cornu and Massot} \cite{cornu79}, which provides only point estimates for CP.
Then, by virtue of the maximum entropy principle \cite{sivia96}, the prior probability distribution of $\tilde{\mathbf C}$ given the tabulated value $c_0$ is an exponential function 
\
\begin{equation}
p(\tilde{c}|I) = p(\tilde{c}|c_0) = \exp (-\lambda \tilde{c})/Z
\quad ,
\end{equation}
\
where $c_0$ is the literature value for the matrix element $\tilde{c}$. Eq. (18) is a general form of the exponential function having the normalization constant $1/Z$ and the scale factor $\lambda$.
The latter can be obtained from the requirement that the expectation value of $\tilde{c}$ on the support $0<\tilde{c}<1$ must be equal to $c_0$
\
\begin{equation}
\langle \tilde{c} \rangle = \int_0^1 d\tilde{c}\ \tilde{c}\ \exp(-\lambda \tilde{c})/Z
= \frac{1}{\lambda} \cdot \frac{1-(1+\lambda) \exp(-\lambda)}{1-\exp(-\lambda)}=c_0
\quad .
\end{equation}
\
For $p(\tilde{\mathbf C}|\mathbf C_0)$ the prior probabilities $p(\tilde{c}|c_0)$ of all elements of the matrix $\tilde{\mathbf C}$ have to be multiplied
\
\begin{equation}
p(\tilde{\mathbf C}|I) = p(\tilde{\mathbf C}|\mathbf C_0) = \prod_{nm} p(\tilde{c}_{nm}|c_{0,nm})
\quad .
\end{equation}
\
Inserting these results into Eq. (13) we have
\
\begin{equation}
\langle \mathbf x_{k} \rangle \sim \int d\tilde{\mathbf C}\ \mathbf x_{k0}\ p(\tilde{\mathbf C} |I)\ \prod_{j} \exp (-\frac{R_j}{2}) | \det \unbfm{Q}_j |^{-\frac12}
\quad .
\end{equation}
\
The integration over $\tilde{\mathbf C}$ is performed by {\it Markov Chain Monte Carlo}(MCMC) method\cite{gilks96} with the sampling density $\rho (\tilde{\mathbf C})$ 
\
\begin{equation}
\rho (\tilde{\mathbf C}) \sim p(\tilde{\mathbf C} |I)\ \prod_{j} \exp (-\frac{R_j}{2}) | \det \unbfm{Q}_j |^{-\frac12}
\quad .
\end{equation}
\
For the second moment of $\mathbf x_k$ we find
\
\begin{equation}
\langle x_{mk}^2 \rangle
=
\int d \tilde{\mathbf C}
\left[
\left(
\mathbf Q_k^{-1}
\right)_{mm}
+
x_{mk0}^2
\right]
\rho ( \tilde{\mathbf C} )
\quad ,
\end{equation}
\
which allows to calculate the variance of $\mathbf x_k$
\
\begin{equation}
\langle \Delta x^2_{mk}\rangle = \langle x^2_{mk}\rangle - \langle x_{mk}\rangle^2
\quad .
\end{equation}
\
Now we have to assign the expectation value and the variance for $\tilde{x}_{Mk}$.
The calculation for $\langle\tilde{x}_{Mk}\rangle$ is straightforward
\
\begin{equation}
\langle \tilde{x}_{Mk}\rangle = \langle 1-\sum_{m=1}^{M-1} \tilde{x}_{mk}\rangle = 1- \sum_{m=1}^{M-1} \langle \tilde{x}_{mk}\rangle 
\quad .
\label{xm1}
\end{equation}
\
Accordingly, for $\langle\tilde{x}_{Mk}^2\rangle$ we have
\
\begin{eqnarray}
\langle \tilde{x}_{Mk}^2\rangle &=& \langle (1-\sum_{m=1}^{M-1} \tilde{x}_{mk}) (1-\sum_{l=1}^{M-1} \tilde{x}_{lk})\rangle \nonumber\\
&=& 1- 2\sum_{m=1}^{M-1} \langle\tilde{x}_{mk}\rangle + \sum_{m,l}^{M-1} \langle\tilde{x}_{mk} \tilde{x}_{lk}\rangle
\quad .
\label{xm2}
\end{eqnarray}
\
From Eq. (25) and (26) it follows
\
\begin{eqnarray}
\langle \Delta\tilde{x}_{Mk}^2\rangle &=& \langle \tilde{x}_{Mk}^2\rangle - \langle \tilde{x}_{Mk}\rangle^2\nonumber\\
&=& 1- 2\sum_{m=1}^{M-1} \langle\tilde{x}_{mk}\rangle + \sum_{m,l}^{M-1} \langle\tilde{x}_{mk} \tilde{x}_{lk}\rangle\\
&& - \left\{1- 2\sum_{m=1}^{M-1} \langle\tilde{x}_{mk}\rangle + \sum_{m,l}^{M-1} \langle\tilde{x}_{mk}\rangle \langle\tilde{x}_{lk}\rangle \right\} \nonumber\\
&=& \sum_{m,l}^{M-1} \left\{ \langle\tilde{x}_{mk} \tilde{x}_{lk}\rangle - \langle\tilde{x}_{mk}\rangle \langle\tilde{x}_{lk}\rangle \right\} \nonumber
\quad .
\end{eqnarray}
\
The estimation of cracking coefficients proceeds in a similar way.
First, Bayes theorem was applied to rewrite the required posterior probability $p(\tilde{\mathbf C}|\mathbf D, \mathbf S, I)$ in terms of a prior on $\tilde{\mathbf C}$ and the marginal likelihood $p(\mathbf D|\tilde{\mathbf C}, \mathbf S, I)$
\
\begin{equation}
p(\tilde{\mathbf C}|\mathbf D, \mathbf S, I) \sim p(\tilde{\mathbf C}|I)\ p(\mathbf D|\tilde{\mathbf C}, \mathbf S, I)
\quad ,
\end{equation}
\
and by using the marginalization rule and the product rule, the likelihood $p(\mathbf D|\tilde{\mathbf C}, \mathbf S, I)$ becomes
\
\begin{eqnarray}
p(\mathbf D|\tilde{\mathbf C}, \mathbf S, I) &=& \int d\mathbf X\ p(\mathbf D, \mathbf X|\tilde{\mathbf C}, \mathbf S, I)\\
&=& \int d\mathbf X\ p(\mathbf X|I)\ p(\mathbf D|\mathbf X, \tilde{\mathbf C}, \mathbf S, I)
\quad .
\end{eqnarray}
\
Finally, we arrive at
\
\begin{equation}
\langle \tilde{\mathbf c}_m^{\mu} \rangle = \int d\tilde{\mathbf C}\ \tilde{\mathbf c}_m^{\mu}\ \rho(\tilde{\mathbf C})
\quad ,
\end{equation}
\
where $\mu =1,2$ for the first and second moment for $\tilde{\mathbf c}_m$.
The sampling density $\rho (\tilde{\mathbf C})$ is the same as in Eq. (22). As a consequence $\tilde{\mathbf C}$ and $\tilde{\mathbf X}$ can be determined by a same MCMC sampling.

\section{Results and Discussion}

All mass spectra presented in this work were taken with a quadrupole mass spectrometer (Hiden, HAL 201), which has the standard axial ion source and the secondary electron multiplier (SEM) for ion detection.
The mass spectrometer chamber with a base pressure better than $1\cdot 10^{-9}$hPa is differentially pumped and connected with the main chamber by a capillary ($\phi$ 1 mm, 10 mm long).
Both chambers are pumped with a separate turbo molecular pump.
The gas flows into the main chamber are controlled by three individual MKS flow controllers (type 1259C), which allow a nominal error of 0.8 \%. During calibration as well as mixture measurements the pressure in the mass spectrometer chamber was kept at about $5\cdot 10^{-8}$hPa to minimize any pressure effect on the cracking patterns.

Fig. 1 depicts the mass signals of 6 mixtures of different gas compositions composed of ethane, propane and n-butane for the most intensive 14 mass channels.
Note the strong similarity of the six spectra.
The quoted ratio was determined from the gas flows into the main chamber and does not reflect the actual concentrations, since the gas flow through the capillary as well as the pumping power in the main and the mass spectrometer chamber strongly depend on species.
However, they still may serve as a crude estimate of the mixture ratio.

The upper part in Fig. 2a shows calibration measurements of C$_2$H$_6$, C$_3$H$_8$ and C$_4$H$_{10}$ together with their literature values taken from {\it Cornu and Massot} \cite{cornu79}. The cracking coefficients are normalized with respect to the sum of the intensities, so that they all have the total intensity of one making the direct comparison of cracking patterns from different sources possible. Although the rough structures of CP from literature and calibration measurements for the three gases look very similar, nearly every peak shows a clear deviation between calibration and literature values.
This is actually expected since the CP depends on the geometry of ion sources, analyzer transmission and mass dependent secondary electron multiplier gain reflecting the specification of a particular mass spectrometer.
The lower plots in Fig. 2a show the results of the Bayesian analysis on the mixture data in Fig. 1.
The agreement of CP between calibration and our analysis is very good. 
For ethane the Bayesian results coincide exactly with those from the calibration, and for propane and butane only few points show a small deviation, namely peaks at m=28 and m=29, 43, respectively. 
But this deviation means not necessarily an error in the measurements or analysis. 
In contrast, it is conceivable that CP of pure gases is changed in a mixture by an influence of gases on each other through a simultaneous presence in the ion source or by variation of partial pressures \cite{breth82}.

Fig. 2b shows the concentrations of the components in six mixtures from the Bayesian analysis and from the independent estimation by pressure measurement. 
For the latter the partial pressure of known gas flows of pure gases was measured by a baratron gauge, which is mounted in the main chamber, and the flux rate of a mixture was rescaled to partial pressures of each gas.
The error of this estimation was assessed to 10 \%, which is mainly contributed by the long-term fluctuation of the total pressure and the tolerance of the flow controller. 
Our results in Fig. 2b are in a satisfactory agreement with the estimation by pressure measurement and 
rather accurate with about 3 \% confidence interval.
It can also be found that there is a systematic deviation between the two sources: the concentrations from the pressure estimation for ethane are lower than those from our analysis and the concentrations for butane show exactly the opposite.
This may be explained by the mass dependent conductance through the capillary \cite{reif65}, which connects the mass spectrometer and the main chamber with the gas inlet system. 
Since the conductance of the capillary is larger for ethane than for butane, the estimations by the baratron have to be increased for ethane and decreased for butane, which leads to a better agreement between the two determinations.
In fact, the determination of the composition of a mixture is rather difficult and needs much care.
A gaseous molecule in a mixture can affect the molecular kinetic of other gases and thus the pressure reading of pure gases cannot be directly combined with the mixture composition.
If no calibration of pure gases is available as in the case of radicals, the pressure determination is even impossible.  Our method in this case provides a unique way to yield concentrations of all components in the mixture.

The accuracy of Bayesian analysis relies on the amount of information provided, which include prior knowledge, number of measurements or the availability of calibration measurements.
It is therefore interesting and important to ask how much information or how many data sets we need to obtain reliable results from the equation system (7).
A single mixture delivers 13 known and 2 unknown parameters ($14-1$ for $\mathbf d_j$ and $3-1$ for $\mathbf x_j$ due to the normalization), while the number of unknown cracking coefficients (=30) is unchanged.
For the latter mass numbers with no significant signal compared to the background noise are neglected.
For the analysis with 3 mixture data, for example, the number of known and unknown parameters are 39 and 36, respectively, illustrating an overdetermined equation system.
This consideration, however, is only valid if the input data are linearly independent. 
Our data by no means meet this condition, as can be seen qualitatively in Fig. 1.
Quantitatively, the significant singular values of the SVD of the data matrix gives the number of sources, which effectively contribute to solving the equation system.
For the data matrix with six column vectors only three singular values are significant (Table 1) indicating that at least six data sets would be needed for the reliable decomposition of the mixtures by the purely mathematical treatment.
In contrast, for the Bayesian approach each data set contributes to increase the accuracy of the analysis on the basis of prior knowledge.
As can be seen in Fig. 2, six mixtures were proved to contain enough information to decompose the mixtures.
Let us now reduce the input data down to three mixtures.
In Fig. 3 the results from three data sets are compared to those from six data sets.
The point estimations of the cracking coefficients are nearly unchanged, but the error bars became somewhat broader, which seems to be reasonable because of less information.
The concentrations derived from 3 and 6 mixtures also show no significant change.
These very satisfactory outcomes from three mixtures would not be possible by directly resolving the equation system (7). 
It should be also noted that the analysis with two mixture data illustrating a superior number of unknown parameters leads to a strong uncertainty in the estimation of both cracking coefficients and concentrations.

We can also even increase the accuracy by adding calibration measurements to our analysis.
Fig. 4 shows the results deduced from 7 data sets including the 6 mixture data and the calibration measurement for butane. 
One can see that the deviation of cracking coefficients at m=29, 43 for butane shown in Fig. 2a entirely vanishes and that the overall error significantly decreases.
This demonstrates that the incorporation of calibration measurements is an efficient way to improve the accuracy of our analysis.
This may be possible even for very unknown mixtures, since in most cases one can infer from the source of mixture gases one or more of its components.

\section{Summary}
We introduced a method for decomposition of multicomponent mass spectra using Bayesian probability theory, which was used to derive concentrations and cracking coefficients of mixture components as well as their confidence interval without any use of calibration measurements.
For synthesized mixtures from ethane, propane and butane the algorithm was able to deliver precise results, which were compared with independent estimations.
This example exhibits a very challenging case, in which the mass spectra of the mixtures show only a moderate difference from each other.
Since the number of available mixture data does not restrict the application of our method but merely affects the accuracy of the analysis, even cases in which limited number of mixtures is available can be processed by our Bayesian approach.
For unknown mixtures, if only few components are assumed to be known from the source of the mixtures, the outcome of our analysis for the concentrations serves as a critical criterion to decide whether a certain species is among the constituents of the mixture.

Future applications include the decomposition of neutral particle fluxes from low temperature hydrocarbon plasmas including larger radicals employed in thin-film deposition and etching. 
The radicals are believed to play an crucial role in film formation at the plasma-surface boundary \cite{jacob98,sugai90}.
Their quantification, which will be possible by the present method, is therefore prerequisite for understanding of the microscopic film growth process.
In this case mixture data are easily available, e.g. by variation of plasma parameters \cite{jacob98}.

\newpage
\bibliographystyle{natbib}

{\bf References}\\[-18mm]

\newpage
{Table 1:Eigenvalues of the singular value decomposition (SVD) of the data matrix with six column vectors shown in Fig. 1.}\\[5mm]

{Fig.1: Mass signals of six different gas mixtures composed of ethane(C$_2$H$_6$), propane(C$_3$H$_8$) and n-butane(C$_4$H$_{10}$). The mixture rates given for C$_2$H$_6$:C$_3$H$_8$:C$_4$H$_{10}$ were determined from gas flows through the gas-inlet device.}\\[5mm]

{Fig.2: (a) upper: cracking patterns from calibration measurements and literature taken from Ref. 6. The mass signals were normalized to the sum of the intensities. lower: Calibration measurements vs. Bayesian analysis using the mixture data in Fig. 1. (b) component concentration of the 6 mixtures from pressure measurement(grey) and from the Bayesian analysis(black). E, P and B in x-axis stand for ethane, propane and butane, respectively. See text for error determination.}\\[5mm]

{Fig.3: Cracking patterns (a) and concentrations (b) of mixture components from Bayesian analysis using 3(black) and 6(grey) mixtures.}\\[5mm]

{Fig.4: Cracking patterns from calibration measurements and Bayesian analysis using 6 mixtures and the calibration measurement of butane.}

\newpage
\vspace{3cm}
\begin{table}[h]
  \begin{center}
    \begin{tabular}{p{15mm}p{15mm}p{15mm}p{15mm}p{15mm}p{15mm}}
      \hline 
      0.9889& 0.1126& 0.0716& 0.0014& 0.0007& 0.0006\\ \hline
    \end{tabular}
  \end{center}
  \caption{Eigenvalues of the singular value decomposition (SVD) of the data matrix with six column vectors shown in Fig. 1.}
\end{table}

\newpage
\begin{figure}
  \begin{center}
    \leavevmode
    \epsfig{file=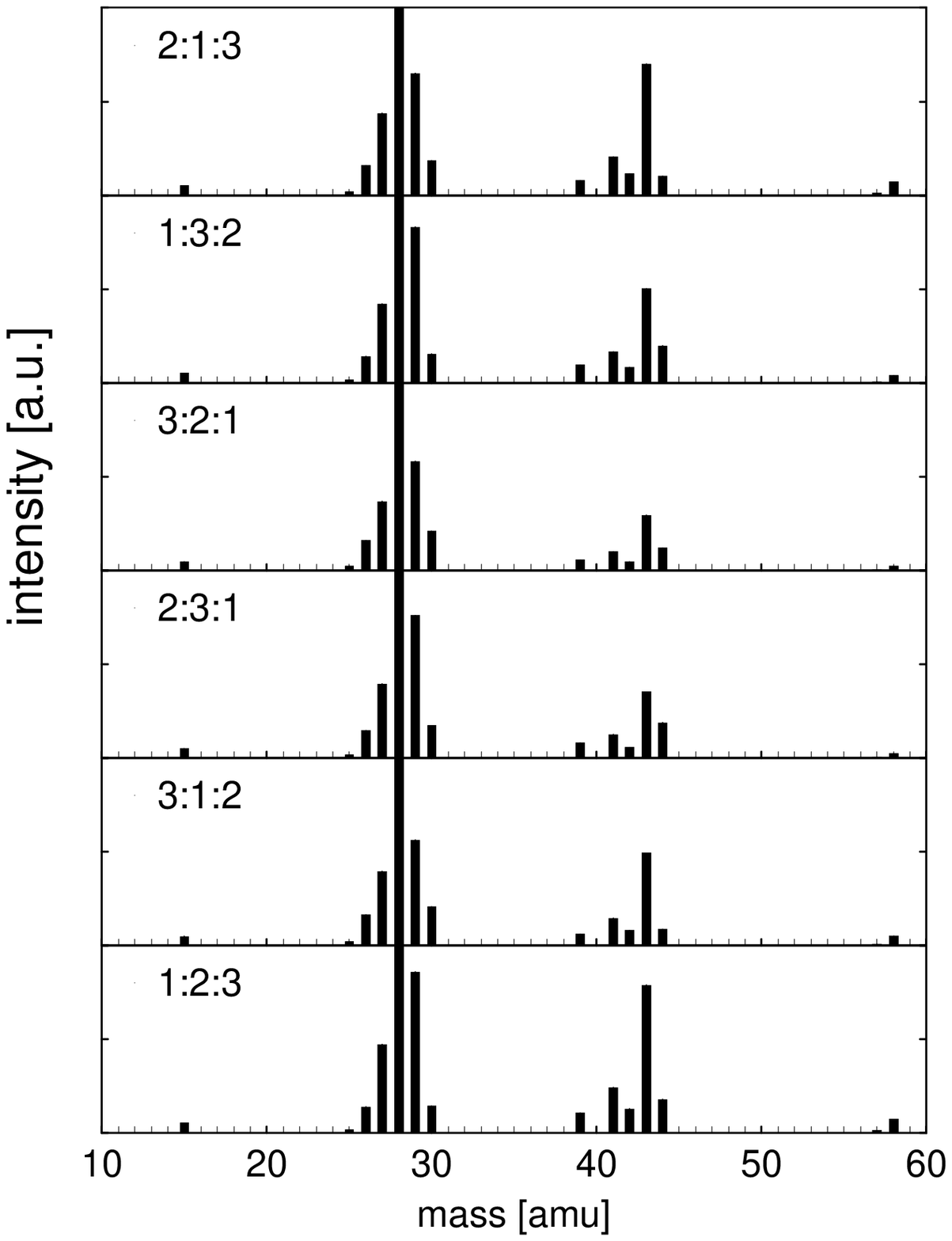,width=130mm}
    \caption{}
  \end{center}
\end{figure}

\begin{figure}
  \begin{center}
    \leavevmode
    \epsfig{file=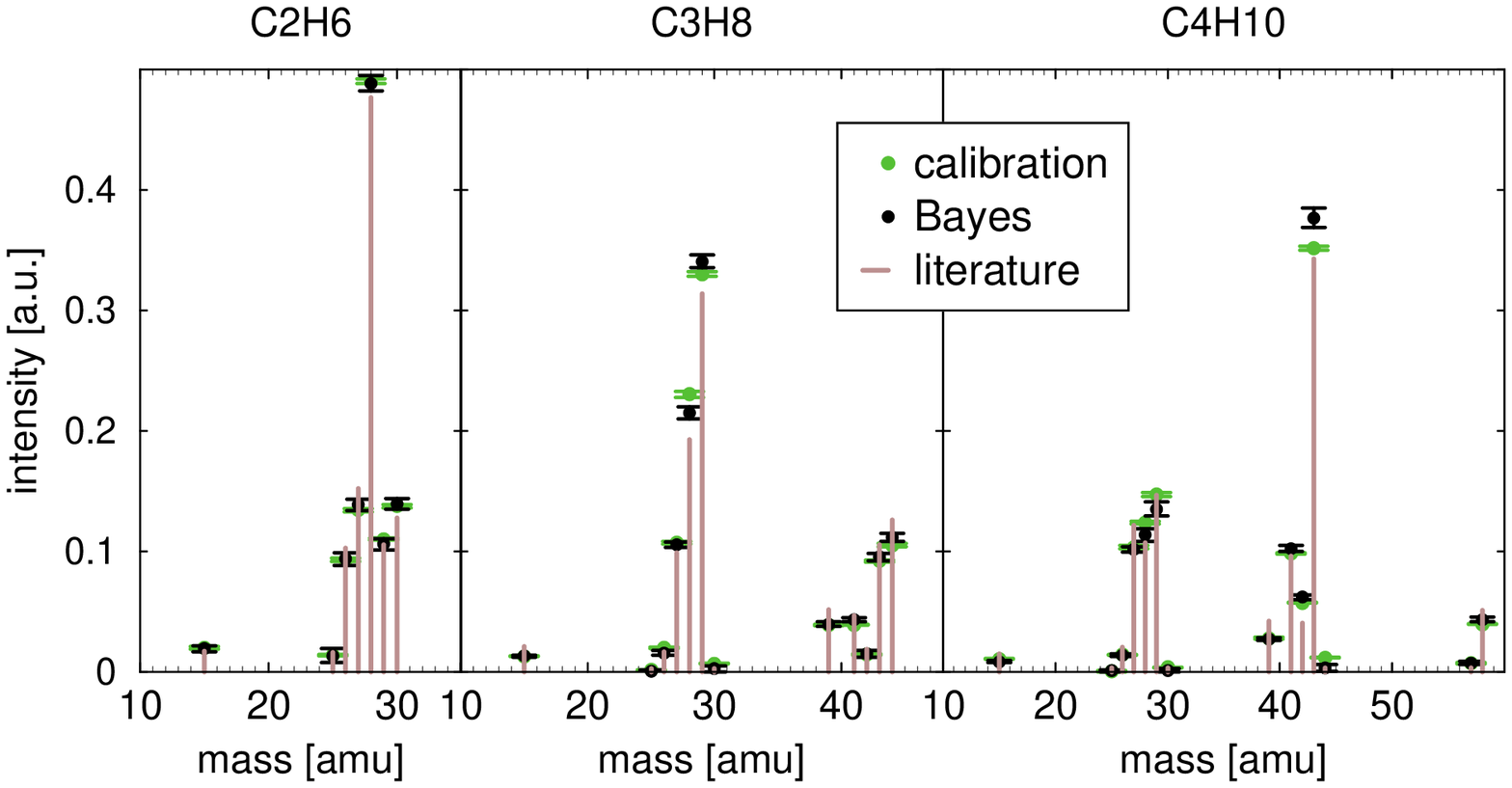,width=13cm,height=75mm}\\[-4mm]
    (a)\\[2mm]
    \epsfig{file=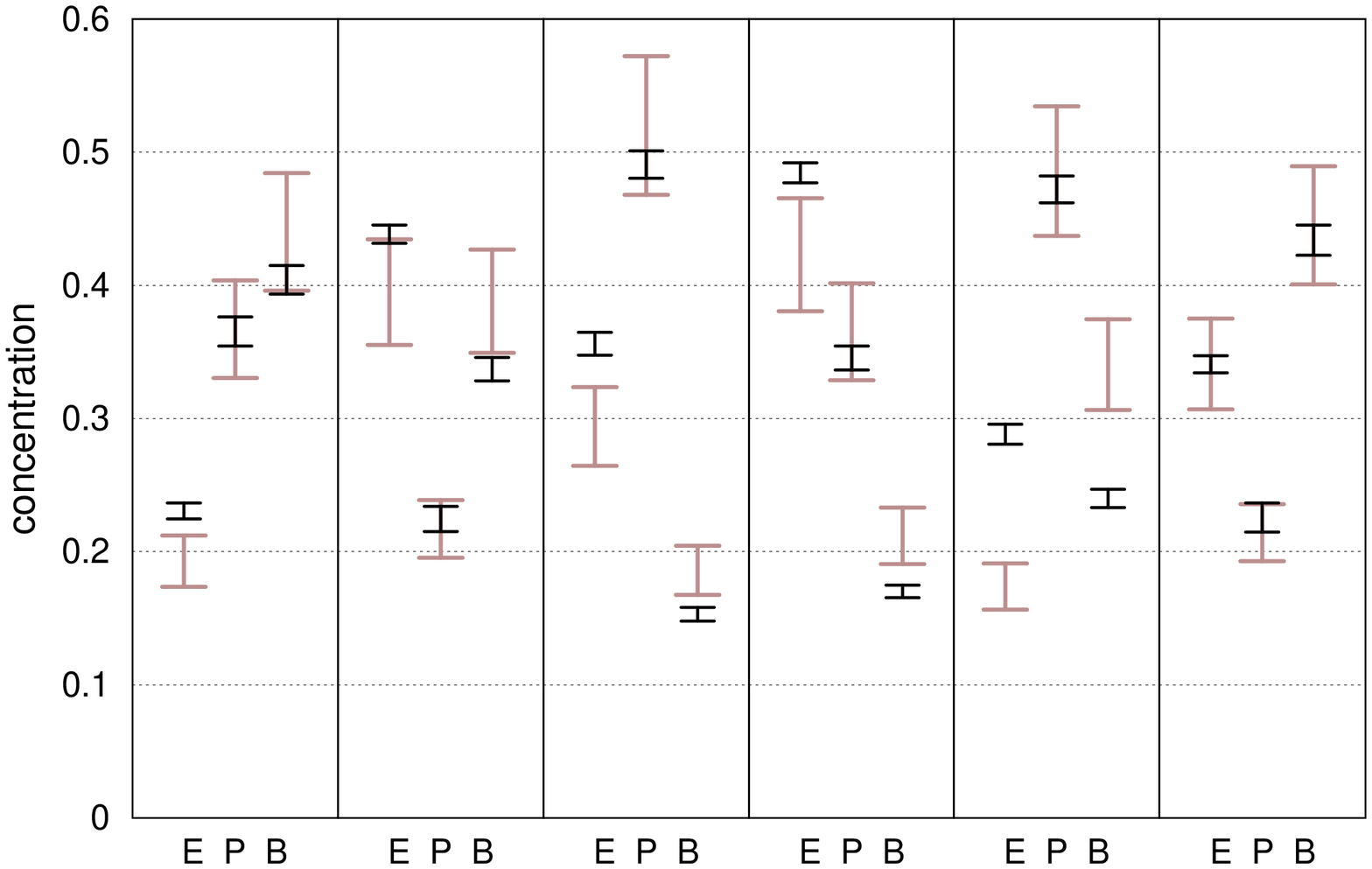,width=13cm,height=80mm}\\[-4mm]
    (b)
    \caption{}
  \end{center}
\end{figure}

\begin{figure}
  \begin{center}
    \leavevmode
    \epsfig{file=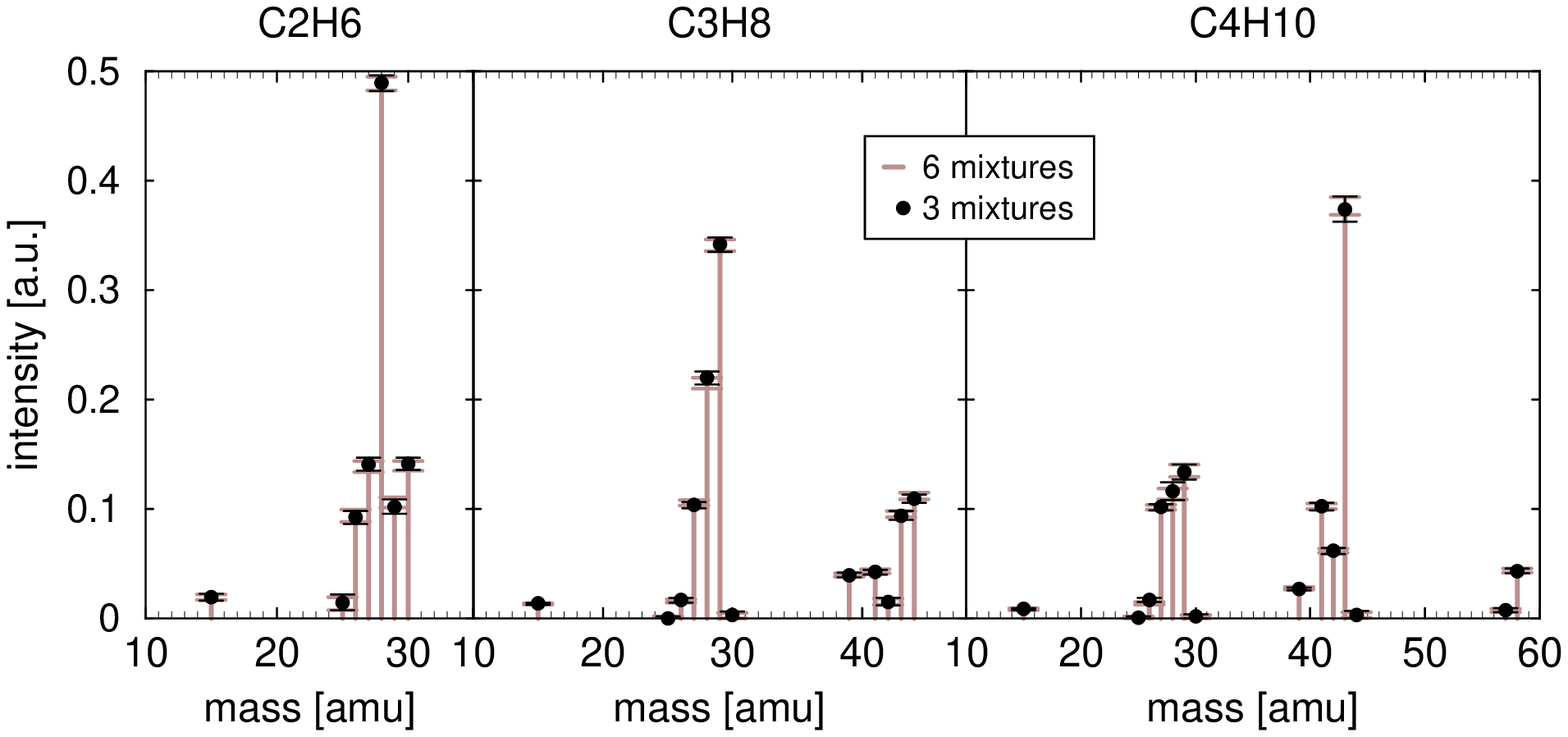,width=130mm,height=75mm}\\
    (a)\\[5mm]
    \epsfig{file=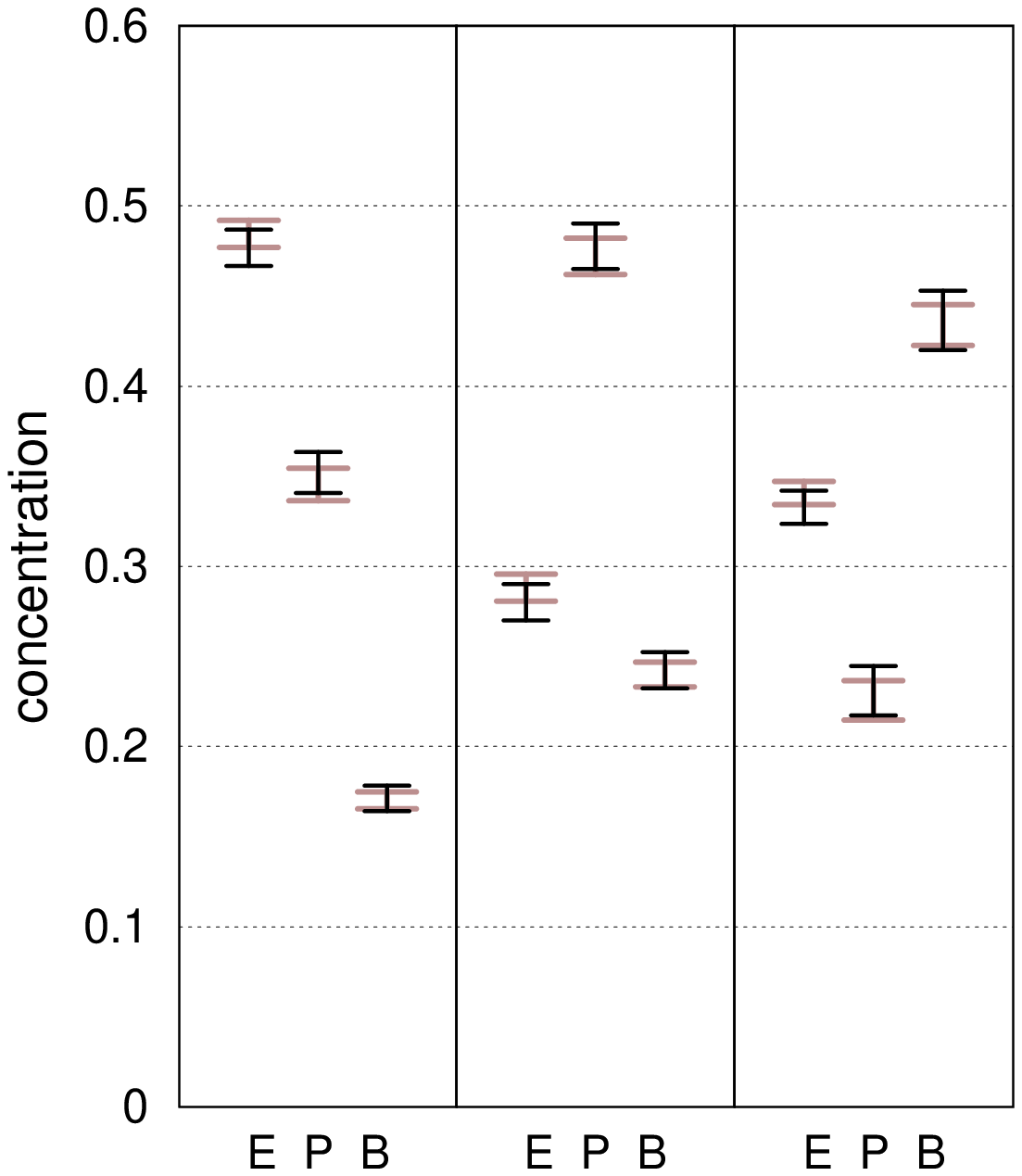,width=90mm,height=90mm}\\
    (b)
    \caption{}
  \end{center}
\vspace{8cm}
\end{figure}

\begin{figure}
  \begin{center}
    \leavevmode
    \epsfig{file=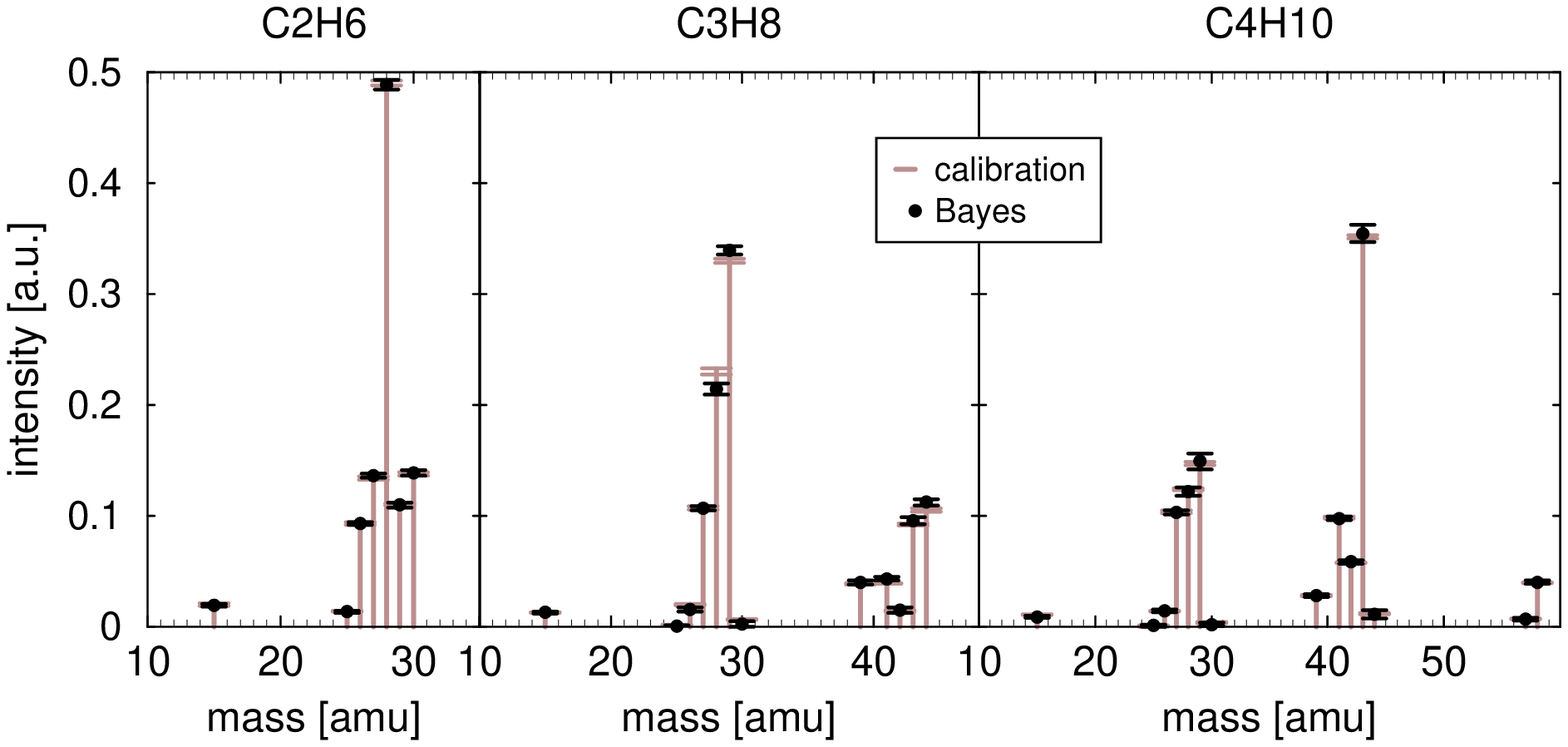,width=130mm,height=75mm}
    \caption{}
  \end{center}
\end{figure}

\end{document}